\documentclass{article}
\usepackage{dcase2019,amsmath,graphicx,url,times,booktabs, tabularx}
\usepackage{kotex, comment, cite}
\usepackage[english,ruled]{algorithm2e}
\usepackage{algpseudocode}
\usepackage{epsfig}
\usepackage{epstopdf}

\title{WEAKLY LABELED SOUND EVENT DETECTION USING TRI-TRAINING AND ADVERSARIAL LEARNING }

 \name{Hyoungwoo Park, Sungrack Yun, Jungyun Eum, Janghoon Cho, Kyuwoong Hwang}
 \address{Qualcomm AI Research$\sthanks{Qualcomm AI Research is an initiative of Qualcomm Technologies, Inc. }$, Qualcomm Korea YH \\
 		$\{$c\_hyoupa, sungrack, c\_jeum, janghoon, kyuwoong$\}$@qti.qualcomm.com}
\begin{document}

\ninept
\maketitle

\begin{sloppy}

\begin{abstract}
This paper considers a semi-supervised learning framework for weakly labeled polyphonic sound event detection problems for the DCASE 2019 challenge's task4 by combining both the tri-training and adversarial learning. The goal of the task4 is to detect onsets and offsets of multiple sound events in a single audio clip. The entire dataset consists of the synthetic data with a strong label (sound event labels with boundaries) and real data with weakly labeled (sound event labels) and unlabeled dataset. Given this dataset, we apply the tri-training where two different classifiers are used to obtain pseudo labels on the weakly labeled and unlabeled dataset, and the final classifier is trained using the strongly labeled dataset and weakly/unlabeled dataset with pseudo labels. Also, we apply the adversarial learning to reduce the domain gap between the real and synthetic dataset. We evaluated our learning framework using the validation set of the task4 dataset, and in the experiments, our learning framework shows a considerable performance improvement over the baseline model.

\end{abstract}


\section{Introduction} 
\label{sec:intro}

The polyphonic sound event detection (SED) has been attracting growing attention in the field of acoustic signal processing\cite{DCASE2016,DCASE2018,7177950,7952260,7933050,7760424,IJCNN,RNN}. The SED aims to detect multiple sound events happened simultaneously as well as the time frame in a sequence of audio events. The applications of the SED include audio event classification\cite{ARaokotomamonjy,DNN2,NAlmaadeed}, media retrieval\cite{MFan,Retrieval} and automatic surveillance\cite{NAlmaadeed} in living environments such as Google Nest Cam\cite{nestcam} which analyzes the audio stream to detect conspicuous sounds such as window breaking and dog barking among various sounds that could occur in daily environments.

Several researches \cite{IJCNN, DNN2,RNN,CNN,CNN2,DCASE2018,7177954,Kumar2016} have been previously proposed . In \cite{7952260}, spectral domain features are used to characterize the audio events, and deep neural networks (DNN)\cite{DNN2} is used to learn a mapping between the features and sound events. In \cite{Kumar2016}, multiple instance learning was exploited to predict the labels of new, unseen instances which rely on an ensemble of instances, rather than individual instances. In \cite{7933050}, convolutional recurrent neural networks (CRNN) was introduced which is a combined network of convolutional neural networks (CNN)\cite{CNN,CNN2} and recurrent neural networks (RNN) \cite{RNN} to get the benefits of both CNN and RNN. In \cite{jiakai2018mean}, Mean Teacher method was adopted where the teacher model is an average of consecutive student models to overcome the limitations of temporal ensembling for semi-supervised learning.



\begin{figure}[t]
	\centering
	\centerline{\epsfig{figure=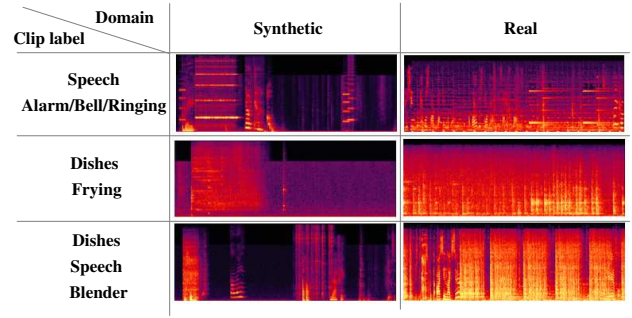,height=4.2cm}}
	\caption{The spectrograms of synthetic and real dataset samples which have the same clip labels}
	\label{fig:intro}
\end{figure}


In contrast to the task 4 of the last year's challenge \cite{dcase2018challenge}, a synthetic dataset with strong annotation is additionally provided in DCASE 2019 challenge's task 4.  Strong annotation includes onset, offset and class label of the sound events. Thus, how to utilize strongly-labeled synthetic data and mutual complement between real dataset and synthetic dataset is a challenging problem in weakly labeled SED problem in DCASE 2019 challenge's task 4. Previous methods have not focused on complement of strongly labeled synthetic dataeset. As shown in Figure \ref{fig:intro}, the log-mel spectrogram of samples which have the same clip label from synthetic data and real data seem too much different. For this reason, we assume that domain gap between synthetic data and real data exists and it causes degradation of performance on test samples. 

This paper presents a sound event detection combining adversarial learning and tri-training. Adversarial learning helps to reduce the gap between synthetic and real data by learning domain-invariant feature while tri-training method \cite{tri-training} which is one of the semi-superived learning methods learns discriminative representations by pseudo labeling one the weakly labeled or unlabeled samples. Pseudo labels are obtained by agreement of output from  confident two labelers on unlable data. Inspired by these properties, we present a weakly labeled polyphonic SED by considering both adversarial learning and tri-training.  

The proposed learning framework was evaluated using a validation set of the DCASE 2019 challenge's task 4 \cite{dcase2019challenge}. In the evaluation results, combined adversarial training and tri-training shows a considerable performance improvement over the baseline model.

\begin{figure*}[t!]
	\centering
	\centerline{\includegraphics[width=1.0\textwidth]{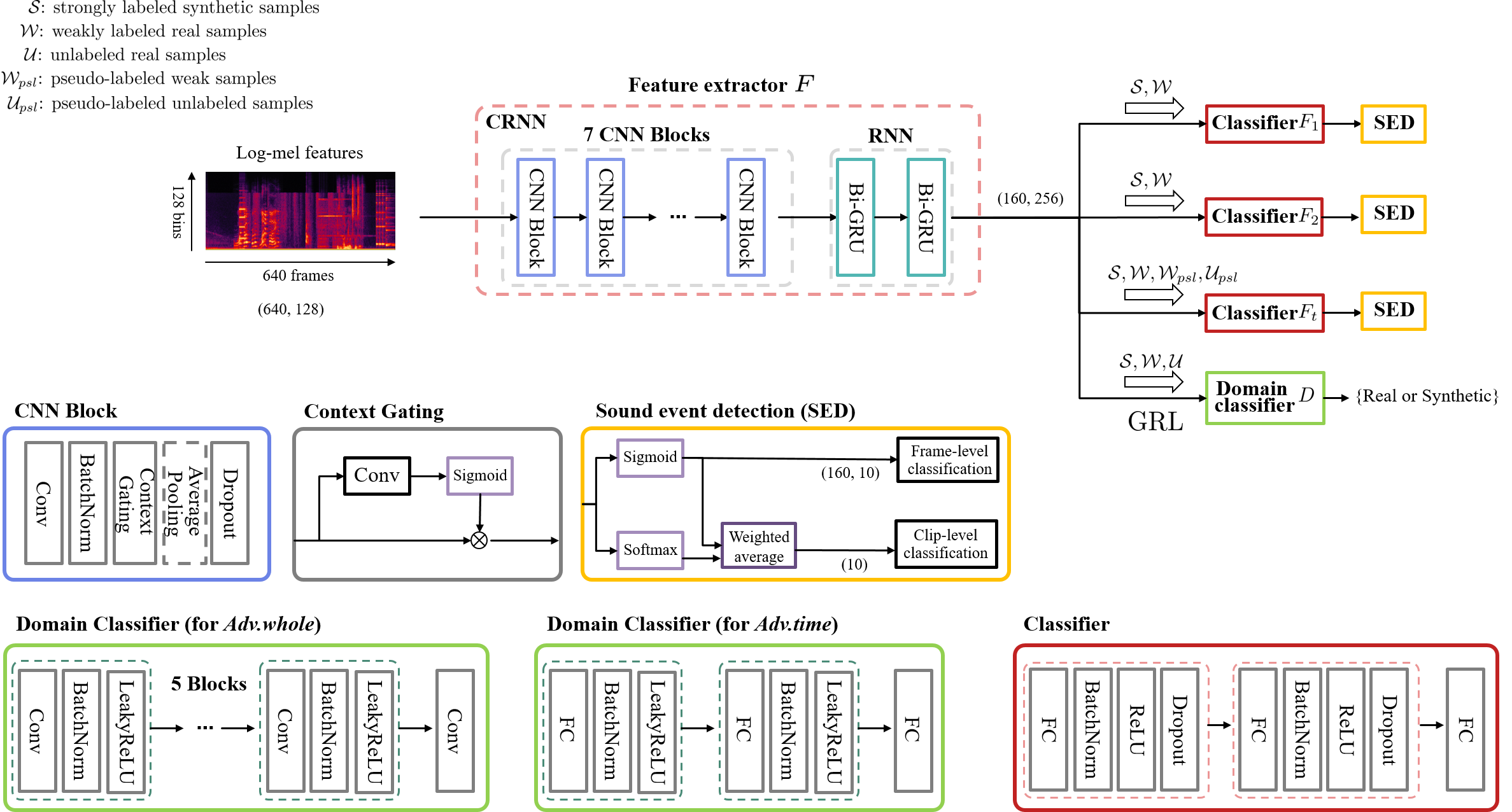}}
	\caption{The proposed learning framework includes feature extractor $F$, classifiers(pseudo-labelers) $F_1, F_2$, final classifier $F_t$ and domain classifier $D$. The dataset to train each component is shown in the figure (e.g. classifier $F_2$ is trained using the strongly-labeled synthetic samples $\mathcal{S}$ and weakly-labeled real samples $\mathcal{W}$. The pseudo-labels are obtained by agreement from two different classifiers $F_1, F_2$ and used in training the final classifier $F_t$. The domain classifier $D$, connected to $F$ via a GRL, classifies the input feature into real or synthetic. With the GRL from $D$ to $F$, the feature distributions between synthetic and real domain become similar, and thus we can obtain the domain-invariant features.} 
	\label{fig:tri_adv}
\end{figure*}


\section{Problem statement and notations}
\label{sec:notations}
For SED, we denote a sound clip by $x\in \mathcal{X}$ and corresponding $y\in \mathcal{Y}$. The SED systems are expected to produce strongly labeled output $y^s$ (i.e. sound class label with start time and end time) from input $x$. However, for weakly labeled SED with semi-supervised setting, dataset consists of strongly labeled data $\mathcal{S} = \{(x_i^s, y_i^s)\}_{i=1}^{m}$, weakly labeled data $\mathcal{W}=\{(x_j^w, y_j^w)\}_{j=1}^{n}$ and unlabeled data $\mathcal{U}=\{x^u_k\}_{k=1}^l$. The weakly labeled data does not provide a temporal range of events but sound class labels detected in a clip. We focus on the usage of weakly labeled or unlabeled data and reducing domain gap between synthetic and real data. Thus, we combine the adversarial learning based on gradient reversal layer (GRL) \cite{ganin2015dann} for reducing the domain gap and tri-training method for pseudo-labeling weakly labeled or unlabeled data such that the networks are learned to output discriminative representations on a real dataset.

\section{Proposed method}
\label{sec:method}
Our proposed method is based on \textit{CRNN} \cite{jiakai2018mean} model, which showed the first place of the task 4 in the last year's challenge by combining with Mean Teacher algorithm \cite{tarvainen2017mean}. The whole architecture is shown in Figure \ref{fig:tri_adv}. A feature extractor $F$, which cosists of seven CNN blocks and two bi-directional gated recurrent units (Bi-GRU) \cite{chung2015gated}, outputs shared features from log-mel features used as input for four networks. Two labelers $F_1$, $F_2$ and classifier $F_t$ predict multiple classes for each time frame and class events for a clip from features extracted by $F$. Let $L_y$ be the classification loss with frame-level classification loss $L_{frame}$ and clip-level classification loss $L_{clip}$ for multi-label prediction.

\begin{equation}
L_{y} = L_{frame} + L_{clip}
\end{equation}

For training with frame-level classification loss and clip-level classfication loss, binary cross entropy (BCE) loss is used with the sigmoid output:
\begin{equation}
L_{clip} = \sum_{i=1}^{N}\sum_{k=1}^{K}[y_{i,k} \log{\hat{y}_{i,k}} + (1-y_{i,k}) \log{(1-\hat{y}_{i,k})}] 
\end{equation}
\begin{equation}
L_{frame} = \sum_{i=1}^{N}\sum_{t=1}^{T}\sum_{k=1}^{K}[y_{i,t,k} \log{\hat{y}_{i,t,k}} + (1-y_{i,t,k}) \log{(1-\hat{y}_{i,t,k})}]
\end{equation}
where $y_{i,k}, y^{i}_{i,t,k} \in [0, 1]$ are the label of sound class $k$ of clip $i$ and the label sound class $k$ at time frame $t$ of clip $i$, respectively. Also, $\hat{y}_{i,k}$ is the predicted probability of sound class $k$ of clip $i$, and $\hat{y}_{i,t,k}$ is $\hat{y}_{i,k}$ at time frame $t$. A domain classifier $D$ classifies features from $F$ into real or synthetic. Based on this architecture, the proposed adversarial learning with tri-training framework for SED will be explained in the next section.

\begin{figure}[t!]
	\centering
	\centerline{\includegraphics[width=0.5\textwidth]{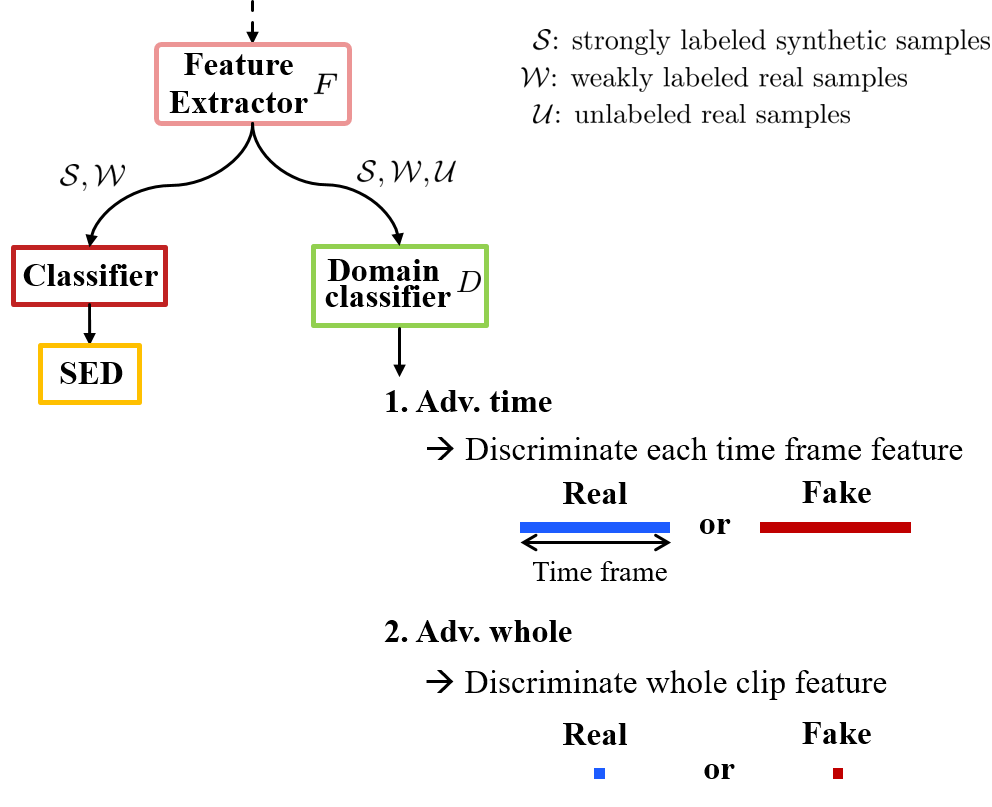}}
	\caption{Two approaches of adversarial learning for sound event detection problems}
	\label{fig:adv}
\end{figure}

\subsection{Adversarial learning}
We denote strongly labeled synthetic dataset by $\mathcal{S}$ and weakly labeled or unlabeled real dataset $\mathcal{W}, \mathcal{U}$ are the different domain (synthetic or real). As shown in Figure 1, since the domain gap between the synthetic and real dataset is quite big, we construct a domain classifier to reduce the gap between two domains by adversarial learning.
The domain classifier $D$ classifies input features into real or synthetic. By applying the GRL\cite{ganin2015dann} from D to the feature extractor F, we can obtain the feature representation whose distributions are almost similar in both real and synthetic domain. We consider two approaches to apply the adversarial learning for SED as shown in Figure \ref{fig:adv}. First, $D$ classifies the whole feature from F into one result: real or synthetic (\textit{Adv.whole}). In this case, GRL makes the features from $F$ domain-invariant. Second, $D$ classifies each time frame of the feature into real or synthetic (\textit{Adv.time}). The second approach is more appropriate than the first one since our architecture predicts multiple sound event classes in each time frame from features extracted from $F$. We denote $\theta_{F}, \theta_{F_1}, \theta_{F_2}, \theta_{F_t}$ and $\theta_{D}$ by the parameters of each network, respectively. Also, $L_d$ is the loss for the domain classification. For training with domain classfication loss, BCE loss is used with the sigmoid output:
\begin{equation}
L_{d} = \sum_{i=1}^{N}[d_{i,t} \log{\hat{d}_{i,t}} + (1-d_{i,t}) \log{(1-\hat{d}_{i,t})}]
\end{equation}
where $d_{i,t}$ is the label of real or synthetic at time frame $t$ of clip $i$, and $\hat{d}_{i,t}$ is predicted probability at time frame $t$ of clip $i$.
Based on GRL, the parameters are updated as follows:
\begin{align}
\theta_{F} &\leftarrow \theta_{F} - \mu \big( \frac{\partial L_y}{\partial \theta_{F}} - \alpha \frac{\partial L_d}{\partial \theta_{F}} \big) \\
\theta_{F_1, F_2, F_t} &\leftarrow \theta_{F_1, F_2, F_t} - \mu \frac{\partial L_y}{\partial \theta_{F_1, F_2, F_t}} \\
\theta_{D} &\leftarrow \theta_{D} - \mu \frac{\partial L_d}{\partial \theta_{D}}
\end{align}
where $\mu,\alpha$ are the learning rate and hyperparameter of GRL, respectively.

\subsection{Tri-training}
We apply the tri-training method to train a network using the pseudo-labeled weakly labeled samples $\mathcal{W}_{psl}$ and pseudo-labeled unlabeled samples $\mathcal{U}_{psl}$. The entire procedure of tri-training is shown in Algorithm \ref{algorithm}. First, we train common feature extractor $F$, two labeling networks $F_1$ and $F_2$ , a final classifier $F_t$ and a domain classifier $D$ with labeled samples $\mathcal{S}$, $\mathcal{W}$ and unlabeled samples $\mathcal{U}$. Second, pseudo-labeled samples are obtained by $F_1$ and $F_2$ trained with labeled samples. When the confidences of both networks' outputs exceed the agreement threshold,  the prediction can be considered reliable. We set this threshold to $0.5$ in the experiments. Also, we expect each labeler to obtain different classifiers $F_1$ and $F_2$ given the same training data, we use the following regularization loss: 
\begin{equation}
L = L_y + \lambda \left| \left(\frac{W_{F_1}}{|W_{F_1}|} \right) ^{\top} \left( \frac{W_{F_2}}{|W_{F_2}|} \right) \right|
\end{equation}
where $W_{F_1} $ and $W_{F_2}$ are weights of first layer of two labelers $F_1$ and $F_2$, respectively. We set $\lambda$ to 1.0 based on the validation set. Then, we use both labeled samples $\mathcal{S}, \mathcal{W}$ and pseudo-labeled samples $\mathcal{W}_{psl}, \mathcal{U}_{psl}$ for training $F, F_t$, and $D$. Then, $F$ and $F_t$ will learn from the labeled real dataset.

\begin{algorithm}[t!]
	\label{algorithm}
	\SetAlgoLined
	\vspace{1mm}
	\textbf{Input:} strongly labeled synthetic data $\mathcal{S} = \big\{(x_{i}^s, y_{i}^s)\big\}^{m}_{i=1}$ \\
	weakly labeled real data $\mathcal{W} = \big\{(x_j^w, y_j^w) \big\}^{n}_{j=1}$ \\
	unlabeled real data $\mathcal{U} = \big\{(x_k^u)\big\}^{l}_{k=1}$\\
	pseudo-labeled weakly labeled data $\mathcal{W}_{psl} = \emptyset$ \\
	pseudo-labeled unlabeled data $\mathcal{U}_{psl} = \emptyset$ \\
	
	\For{i = 1 \KwTo iter}{
		Train $F, F_{1}, F_{2}, F_{t}, D $ with mini-batch from labeled training set $ \mathcal{S}, \mathcal{W},\mathcal{U} $ \\
	} 
	$\mathcal{W}_{psl} =$ \textit{Pseudo-labeling}($F, F_{1}, F_{2}, \mathcal{W}$)\\
	$\mathcal{U}_{psl} =$ \textit{Pseudo-labeling}($F, F_{1}, F_{2}, \mathcal{U}$)\\
	
	\For{j =1 \KwTo iter}{
		Train $ F, F_{t}, D$ with mini-batch from labeled training set $ \mathcal{S}, \mathcal{W}$ and pseudo-labeled training set $\mathcal{W}_{psl}, \mathcal{U}_{psl}$ \\
	}
	
	
	\caption{The function \textit{Pseudo-labeling} is the process of assigning pseudo-labeling based on agreement threshold from two labelers. We assign pseudo-labels to weakly labeled or unlabeled samples when both predictions of $F_1$ and $F_2$ are confident and agreed to the same prediction. }
\end{algorithm}

\section{EXPERIMENTS}
\label{sec:experiments}
\subsection{Dataset}

The DCASE 2019 challenge's task 4\cite{dcase2019challenge} provides the following 3 subsets of the dataset in the training: 1,578 clips of the weakly labeled set, 14,412 clips of the unlabeled in-domain set and 2,045 clips of the synthetic set with strong annotations of events and timestamps. Weakly labeled and unlabeled in-domain sets are from Audioset\cite{Audioset} which drawn from 2 million YouTube videos. The synthetic set is generated with Scaper\cite{Scaper} to increase the variability of the output for soundscape synthesis and augmentation. These audio clips are 10 second-long and contain one or multiple sound events among 10 different classes (speech, dog, cat, alarm/bell/ringing, dishes, frying, blender, running water, vacuum cleaner and electric shaver/toothbrush) which may partly overlap. 

\subsection{Experimental setup}

The model was developed using PyTorch \cite{paszke2017automatic} and all experiments were conducted on an a GeForce GTX TITAN X GPU 12GB RAM. Also, our architecture was trained with a mini-batch size of $64$ using Adam optimizer \cite{kingma2014adam} with an initial learning rate of $0.001$ and exponential decay rate for the $1st$ and $2nd$ moments of $0.9$ and $0.999$, respectively. The input audio clips are down-sampled from 44.10 kHz to 22.05 kHz. 
And, the log-mel spectrogram is extracted from the audio clip with the size of $640\times 128$: 128-bin is used, and 2048-window with 345-hop is used to convert into 640 frames.

\subsection{Experimental results}
We evaluated our proposed framework using the DCASE 2019 challenge's task4 validation dataset. We could not measure performance on evaluation dataset since the labels of evaluation dataset are not available yet. The macro event-based F1 scores and segment-based F1 scores on validation dataset are shown in Table \ref{table:table1}. Segment-based metrics evaluate an active/inactive state for each sound event in a fixed-length interval, while event-based metrics evaluate sound event class detected in the fixed-length interval. The baseline of DCASE 2019 challenge's task 4 used the Top-1 ranked model \cite{jiakai2018mean} of DCASE 2018 challenge's task 4, which proposed Mean Teacher method for SED; however, this baseline was designed with smaller architecture. Thus, we designed our baseline model based on originally Top-1 ranked model in DCASE 2018. Our baseline consists of feature extractor $F$ and classifier $F_t$ which are trained using the strongly labeled synthetic data and weakly labeled real data without Mean Teacher algorithm. The baseline showed 24.15$\%$ event-based F1 score. For the comparisons, the official results of various SED frameworks submitted for the DCASE 2019 challenge's task 4 are shown in Table \ref{table:table1}. 

With our baseline model, we applied the adversarial learning method in two ways. The domain classifier $D$ predicted real or synthetic on whole feature map (\textit{Adv.whole}) and on features in each time frame (\textit{Adv.time}). Both adversarial learning approaches improved the performance as shown in Table \ref{table:table1}. These approaches reduced domain gap between synthetic and real feature distributions, thsu we could improve performance since the validation set was also from the real audio clips. The \textit{Adv.time} and \textit{Adv.whole} achieved 31.33$\%$ and 30.65$\%$, respectively. The \textit{Adv.time} method showed better performance than the \textit{Adv.whole} since the architecture tries to predict multi-label in each time frame. We also performed pseudo-labeling method using tri-training procedure. The tri-training method achieved 30.23$\%$. Tri-training method showed better performance than the adversarial learning in evaluating segment-based F1 scores.
Finally, we evaluated the combined architecture: adversarial learning with the tri-training. When we trained two labelers in tri-training, we also trained domain classifier simultaneously for adversarial learning. After training two labelers with adversarial learning, more confident labelers for predicting sound event classes on real dataset were obtained. Then, we assigned pseudo-label to weakly labeled and unlabeled samples based on two labelers. We trained the final classifier with labeled samples and pseudo-labeled samples by tri-training scheme. In combining the adversarial learning with tri-training method, we considered the previous two approaches: \textit{Adv.whole} and \textit{Adv.time}. The tri-training method combined with \textit{Adv.whole} approach showed 32.64$\%$ event-based F1 score and the tri-training method combined with \textit{Adv.time} achieved 35.10$\%$. \textit{Adv.time}+Tri-training achieved the highest event-based F1 score of our models. The tri-training method achieved 62.86$\%$ segment-based F1 score and it is better than \textit{Adv.time}+Tri-training. We think that adversarial learning contributes more to inference exact sound label in time frame while the tri-training contributes more to inference the exact boundary of the sound event.

\section{CONCLUSION}
\label{sec:conclusion}
In this paper, we consider the semi-supervised learning framework for weakly labeled SED problem for the DCASE 2019 challenge's task4 by combining both tri-training and adversarial learning. The entire dataset consists of the synthetic data with the strong label (sound event labels with boundaries) and real data with weakly labeled (sound event label) and unlabeled dataset. We reduce domain gap between strongly labeled synthetic dataset and weakly labeled or unlabeled real dataset to train networks to learn domain-invariant feature for preventing degradation of performance. Also, we utilize pseudo labeled samples based on confident multiple labelers trained by labeled samples. Then, networks learn the discriminative representation of the unlabeled dataset. The tri-training method combined with adversarial learning on each time frame shows a considerable performance improvement over the baseline model. 

\begin{table}[]
	\centering
	\caption{The event based macro F1 scores and segment based macro F1 scores of proposed methods on validation dataset in DCASE 2019 challenge's task 4}
	\begin{tabular}{lcc}
		\hline
		\textbf{Model}     & \multicolumn{2}{c}{\textbf{Macro F1 (\%)}}       \\
		& \multicolumn{1}{c}{Event-based} & \multicolumn{1}{c}{Segment-based} \\ \hline
		
		Wang\_YSU\_task4\_1 	& 19.4$\%$					   & - 		\\
		Kong\_SURREY\_task4\_1  & 21.3$\%$					   & -		\\
		Wang\_NUDT\_task4\_3 	& 22.4$\%$					   & - 		\\
		DCASE 2019 baseline \cite{dcase2019challenge}&23.7$\%$ &    55.2$\%$  \\
		Rakowski\_SRPOL\_task4\_1 & 24.3$\%$				   & -		\\
		mishima\_NEC\_task4\_4 & 24.7$\%$					   & -		\\
		Lee\_KNU\_task4\_3     & 26.7$\%$					   & - 		\\
		bolun\_NWPU\_taks4\_2  & 31.9$\%$					   & - 		\\
		Kothinti\_JHU\_task4\_1& 34.6$\%$					   & -		\\
		ZYL\_UESTC\_task4\_2   & 35.6$\%$					   & -		\\
		Kiyokawa\_NEC\_task4\_4& 36.1 $\%$					   & - 		\\
		PELLEGRINI\_IRIT\_task4\_1 & 39.9$\%$                  & -		\\
		Lim\_ETRI\_task4\_4    & 40.9 $\%$					   & - 		\\
		Shi\_FRDC\_task4\_2    & 42.5$\%$                      & -				\\ 
		Yan\_USTC\_task4\_4    & 42.6 $\%$                     & -		\\
		Delphin\_OL\_task4\_2  & 43.6$\%$                      & -		\\
		Lin\_ICT\_task4\_3     & \textbf{45.3$\%$}             & -		\\ \hline
		Our baseline           & 24.15$\%$                     & 57.70$\%$    \\
		\textit{Adv.whole}     & 30.65$\%$                     & 59.06$\%$    \\
		\textit{Adv.time}      & 31.33$\%$                     & 59.26$\%$    \\
		Tri-training           & 30.23$\%$                     & \textbf{62.86$\%$} \\
		\textit{Adv.whole} + Tri-training & 32.64$\%$          & 60.48$\%$    \\
		\textit{Adv.time} + Tri-training  & \textbf{35.10$\%$} & 60.67$\%$      \\ \hline
	\end{tabular}
	\label{table:table1}
\end{table}


\newpage
\bibliographystyle{IEEEtran}
\bibliography{refs}
%
%
%
%
%
%
%
%
%

\end{sloppy}
\end{document}